\documentclass[a4paper,11pt]{article}
\pdfoutput=1
\usepackage{etex}
\usepackage{graphicx,comment,booktabs}
\usepackage[table]{xcolor}
\usepackage[top=2.5cm,bottom=2.5cm,left=2.5cm,right=2.5cm]{geometry}
\usepackage{amsmath,amssymb,bm,framed,natbib,authblk}

\usepackage{xcolor}
\definecolor{dgreen}{rgb}{0,0.6,0} \definecolor{dred}{rgb}{0.6,0,0}
\definecolor{dpurple}{HTML}{A020F0} \definecolor{dblue}{rgb}{0,0,1}
\definecolor{hlcolor}{rgb}{1,1,0.8}

\usepackage{lineno}

\usepackage{soul}
\sethlcolor{hlcolor}

\usepackage[colorlinks=true,linkcolor=dblue,citecolor=dred]{hyperref}
\usepackage{cleveref}
\creflabelformat{equation}{#2#1#3}
 
\title{Characterizing variability in nonlinear,\\ recurrent neuronal networks}

\author[1]{Guillaume Hennequin${}^{\rm @}$} 
\author[1]{M\'at\'e Lengyel}

\affil[1]{Computational and Biological Learning Lab, Department of Engineering,
University of Cambridge, Cambridge, UK}

\date{${}^{\rm @}$ g.hennequin@eng.cam.ac.uk \\ \vspace*{1em} \today}

\newcommand\s[1]{\tilde{#1}}
\renewcommand\b\mathbf
\renewcommand\d{\mathrm{d}}
\newcommand\e{\mathrm{E}}
\renewcommand\i{\mathrm{I}}
\newcommand\tm{\tau}
\newcommand\sig[1]{\sqrt{\Sigma_{#1 #1}}}

\newcommand\trans{{\mathop{\rm T}}}
\newcommand\weff{\bm{\mathcal{J}}}
\newcommand\diag{\text{diag}}

\begin{document}

\maketitle

%<<<1 Abstract
\begin{abstract} In this note, we develop semi-analytical techniques to obtain
the full correlational structure of a stochastic network of nonlinear neurons
described by rate variables. Under the assumption that pairs of membrane
potentials are jointly Gaussian -- which they tend to be in large networks --
we obtain deterministic equations for the temporal evolution of the mean firing
rates and the noise covariance matrix that can be solved straightforwardly
given the network connectivity. We also obtain spike count statistics such as
Fano factors and pairwise correlations, assuming doubly-stochastic action
potential firing.  Importantly, our theory does not require fluctuations to be
small, and works for several biologically motivated, convex single-neuron
nonlinearities.  \end{abstract}

\tableofcontents

%<<<1 Introduction

% ================================================================================
% === Document                                                                 ===
% ================================================================================
 
\section{\label{sec:intro}Introduction}

In this technical note, we develop a novel theoretical framework for characterising
across-trial (and temporal) variability in stochastic, nonlinear recurrent
neuronal networks with rate-based dynamics. We consider networks in which
momentary firing rates $\b{r}$ are given by a nonlinear function of the
underlying (coarse-grained) membrane potentials $\b{u}$. In particular, we
treat the case of the threshold-power-law nonlinearity $r=k \lfloor u
\rfloor_+^n$ (where $n$ is an arbitrary positive integer), which has been shown
to approximate the input-output function of real cortical neurons (with $n$
ranging from 1 to 5; \citealp{Priebe04,Miller02}). This model is of particular
interest as it captures many of the nonlinearities observed in the
trial-averaged responses of primary visual cortex (V1) neurons to visual
stimuli \citep{Ahmadian13,Rubin15}, as well as the stimulus-induced suppression
of \mbox{(co-)variability} observed in many cortical areas
(\citeauthor*{HennequinSubmitted}, \emph{submitted}).

We derive assumed density filtering equations that describe the temporal
evolution of the first two moments of the joint probability distribution of
membrane potentials and firing rates in the network. These equations are based
on the assumption that pairs of membrane potentials are jointly Gaussian at all
times, which tends to hold in large networks. This approach allows us to solve
the temporal evolution of the mean firing rates and the across-trial noise
covariance matrix, given the model parameters (network connectivity,
feedforward input to the network, and statistics of the input noise). We also
obtain the full firing rate cross-correlogram for any pair of neurons, which
allows us to compute both Fano factors and pairwise spike count correlations in
arbitrary time windows, assuming doubly stochastic (inhomogeneous Poisson)
spike emission. Importantly, our theory does not require fluctuations to be
small, and is therefore applicable to physiologically relevant regimes where
spike count variability is well above Poisson variability (implying large
fluctuations of $\b{u}$, which is therefore often found below the threshold of
the nonlinearity).

This note is structured as follows. We first provide all the derivations, together
with a summary of equations and some practical details on implementation. We
then demonstrate the accuracy of our approach on two different networks of
excitatory and inhibitory neurons: a random, weakly connected network, and a
random, strongly connected but inhibition-stabilized network.  In the latter
case, the stochastic dynamics are dominated by balanced (nonnormal)
amplification, leading to the emergence of strong correlations between
neurons~\citep{Murphy09,Hennequin14} which could in principle make the membrane
potential distribution non-Gaussian -- the only condition that could break the
accuracy of our theory. Even in this regime, we obtain good estimates of the
first and second-order moments of the joint distributions of both $\b{u}$ and
$\b{r}$.

%<<<1 Theoretical results
\section{\label{sec:theory}Theoretical results}

\paragraph{Notations} We use $\cdot^\trans$ to denote the vector/matrix
transpose, bold letters to denote column vectors ($\b{v} \equiv (v_1, \ldots,
v_N)^\trans$) and bold capitals for matrices (e.g.\ $\b{W} = \{W_{ij}\}$) . The
notation $\diag(\b{v})$ is used for the diagonal matrix with entries $v_1, v_2,
\ldots, v_N$ along the diagonal.
 
%<<<2 Model setup
\subsection{Model setup}

We consider a network of $N$ interconnected neurons, whose continuous-time,
stochastic and nonlinear dynamics follow
\begin{equation}\label{eq:dynamics-fast-noise}
\d u_i \quad=\quad \frac{\d t}{\tm_i} \left(-u_i(t) + h_i(t) 
  + \sum_j W_{ij}\: r_j(t)\right) + \d\chi_i
\end{equation}
where the momentary firing rate $r_j(t)$ is given by a positive, nonlinear
function $f$ of $u_j(t)$:
\begin{equation}\label{eq:nonlinearity}
r_j(t) = f\left[u_j(t)\right]
\end{equation}
The first few derivations below hold for arbitrary nonlinearity $f$; we will
commit to the threshold-powerlaw nonlinearity $f(u) \propto \lfloor u
\rfloor_+^n$ only later. In \Cref{eq:dynamics-fast-noise}, $\tm_i$ is the intrinsic
(``membrane'') time constant of unit $i$, $h_i(t)$ denotes a deterministic and
perhaps time-varying feedforward input, $W_{i\ell}$ is the synaptic weight from
unit $\ell$ onto unit $i$, and $\d\boldsymbol\chi$ is a multivariate Wiener
process with covariance $\langle \chi_i(t) \: \chi_j(t+s) \rangle =
\Sigma^\chi_{ij} \delta(s)$. Note that the elements of $\b\Sigma^\chi$ have
units of $({\rm mV})^2/s$. We will later extend \Cref{eq:dynamics-fast-noise} to
the case of temporally correlated input noise.

Most of our results can be derived from the finite-difference version of
\Cref{eq:dynamics-fast-noise}, given for very small $\epsilon$ by
\begin{equation}\label{eq:model} 
u_i(t)-u_i(t-\epsilon) \quad=\quad
\frac{\epsilon}{\tm_i} \left[
  -u_i(t-\epsilon) + h_i(t) + \sum_\ell W_{i\ell}\:r_\ell(t-\epsilon) \right]
  + \sqrt{\epsilon}\:\chi_i(t)
\end{equation}
Here $\boldsymbol\chi(t)$ is drawn from a multivariate normal distribution with
covariance matrix $\b\Sigma^\chi$, independently of the state of the sytem in
the previous time step ($t-\epsilon$).

We now show how to derive deterministic equations of motion for the first and
second-order moments of $\b{u}$, then explain how these can be made
self-consistent using the Gaussian assumption mentioned in the introduction.
Solving these equations is straightforward, and returns the mean membrane
potential of each neuron as well as the full matrix of pairwise covariances.
We also show how the moments of $\b{r}$ can be derived from those of $\b{u}$,
and how Fano factors and spike count correlations can then be obtained from
those.

%<<<2 Temporal evolution of the moments of u
\subsection{\label{sec:umoments}Temporal evolution of the moments of the voltage}

We are interested in the moments of the membrane potential variables, defined as
\begin{subequations}
\begin{align}
\mu_i(t) \quad &\equiv \quad \langle u_i(t) \rangle 
\label{eq:def:mu}\\
\Sigma_{ij}(t,s) \quad &\equiv \quad \langle \s{u}_i(t) \:\s{u}_j(t+s) \rangle
\label{eq:def:sigma}
\end{align}
\end{subequations}
where $\langle\cdot(t)\rangle$ denotes ensemble averaging (or
``trial-averaging'', i.e.\ an average over all possible realisations of the
noise processes $\boldsymbol{\chi}(t')$ for $0\leq t'\leq t$), and with the notation
$\s{z}(t) \equiv z(t) - \langle z(t)\rangle$. Note that
$\langle \cdot(t) \rangle$ will coincide with temporal averages in the stationary case
when the dynamics has a single fixed point (ergodicity). Before we proceed, let
us introduce similar notations for other moments:
\begin{subequations}
\begin{align}
\nu_i(t) \quad&\equiv\quad \langle r_i(t) \rangle \label{eq:def:nu} \\
\Gamma_{ij}(t,s) \quad&\equiv\quad 
  \langle \tilde{u}_i(t) \: \tilde{r}_j(t+s) \rangle \label{eq:def:gamma} \\
\Lambda_{ij}(t,s) \quad&\equiv\quad
  \langle \tilde{r}_i(t)\: \tilde{r}_j(t+s) \rangle \label{eq:def:lambda}
\end{align}
\end{subequations}
Taking the ensemble average of \Cref{eq:model} and the limit $\epsilon\to0$
yields a differential equation for the membrane potential mean:
\begin{equation}
\tm_i\: \frac{\d\mu_i}{\d t} = \label{eq:mean} - \mu_i(t) + h_i(t) 
+ \sum_j W_{ij}\: \nu_j(t)
\end{equation}
We now subtract its average from \Cref{eq:model}:
\begin{equation}\label{eq:finitediff}
\s{u}_i(t)- \s{u}_i(t-\epsilon) = \frac\epsilon{\tm_i} \left[
-\s{u}_i(t-\epsilon) + \sum_j W_{ij}\:\s{r}_j(t-\epsilon)
\right] + \sqrt\epsilon \: \chi_i(t)
\end{equation}
The equations of motion for the variances and covariances can be obtained in a
number of ways. Here we adopt a simple approach based on the finite differences
of \Cref{eq:finitediff}, observing that
\begin{equation}\label{eq:trick}
\s{u}_i(t)\:\s{u}_j(t) - \s{u}_i(t-\epsilon)\:\s{u}_j(t-\epsilon) 
  = 
      \s{u}_i(t)\: \left[\s{u}_j(t) - \s{u}_j(t-\epsilon)\right]
      + \s{u}_j(t-\epsilon) \left[ \s{u}_i(t) - \s{u}_i(t-\epsilon)\right]
\end{equation}
Substituting \Cref{eq:finitediff} into \Cref{eq:trick}, we obtain
\begin{align}\label{eq:rcalc}
\s{u}_i(t)\:\s{u}_j(t) - \s{u}_i(t-\epsilon)\:\s{u}_j(t-\epsilon) =
  &-\frac\epsilon{\tm_i}\: \s{u}_i(t-\epsilon) \: \s{u}_j(t-\epsilon)
    -\frac\epsilon{\tm_j}\: \s{u}_i(t) \: \s{u}_j(t-\epsilon)  \\
  &+\epsilon\sum_\ell \s{r}_\ell(t-\epsilon)
       \left(\frac{W_{i\ell}}{\tm_i}\:\s{u}_j(t-\epsilon)
             + \frac{W_{j\ell}}{\tm_j}\:\s{u}_i(t)\right) \nonumber\\
  &+\sqrt\epsilon \left(
      \chi_i(t)\:\s{u}_j(t) 
    + \chi_j(t)\:\s{u}_i(t-\epsilon)\right) \nonumber
\end{align}
We now take ensemble expectations on both sides. The l.h.s.\ of \Cref{eq:rcalc}
averages to $\Sigma_{ij}(t) - \Sigma_{ij}(t-\epsilon)$. The term $\chi_j(t)\:
\s{u}_i(t-\epsilon)$ averages to zero for any $(i,j)$ pair, because the input
noise terms at time $t$ are independent of the state of the system at time
$t-\epsilon$. However, we expect the equal-time product $\chi_i(t)\:
\s{u}_j(t)$ to average to something small ($\mathcal{O}(\sqrt\epsilon)$) but
non-zero. Let us compute it, by recalling that
\begin{equation} 
\s{u}_j(t) = \s{u}_j(t-\epsilon) 
+ \frac\epsilon{\tm_j}\left[\:\text{something at time}\: (t-\epsilon)\:\right]
+ \sqrt\epsilon\:\chi_j(t)
\end{equation}
Multiplying both sides by $\chi_i(t)$ and taking expectations, again all
products of $\chi_i(t)$ with quantities at time $t-\epsilon$ vanish, and we are
left with 
\begin{equation}
\langle \s{u}_j(t)\:\chi_i(t)\rangle \quad=\quad \Sigma^\chi_{ij} \sqrt\epsilon
\end{equation}
Thus, when averaged, \Cref{eq:rcalc} becomes
\begin{equation}
\frac{\Sigma_{ij}(t) - \Sigma_{ij}(t-\epsilon)}\epsilon = 
  -\frac{\Sigma_{ij}(t-\epsilon,0)}{\tau_i}
  -\frac{\Sigma_{ij}(t,-\epsilon)}{\tau_j}
  +\sum_\ell \left( 
    \frac{W_{i\ell}}{\tau_i} \Gamma_{j\ell}(t-\epsilon,0) +
    \frac{W_{j\ell}}{\tau_j} \Gamma_{i\ell}(t,-\epsilon)
    \right)
  + \Sigma_{ij}^\chi
\end{equation}
Now taking the limit of $\epsilon \to 0$, and by continuity of both $\b{u}$ and
$\b{r}$, we obtain the desired equation of motion of the zero-lag covariances:
\begin{equation}\label{eq:fpcov}
  \frac{\d\Sigma_{ij}(t,0)}{\d t} =
  \Sigma^\chi_{ij}
  + \frac1{\tm_i} \left( -\Sigma_{ij}(t,0) + \sum_\ell W_{i\ell}\: \Gamma_{j\ell}(t,0)\right)
  + \frac1{\tm_j} \left( -\Sigma_{ij}(t,0) + \sum_\ell W_{j\ell}\: \Gamma_{i\ell}(t,0)\right)
\end{equation}

In the special case of \emph{constant} input $h_i(t)$, the covariance matrix of
the steady-state, stationary distribution of potentials satisfies
\Cref{eq:fpcov} with the l.h.s.\ set to zero. This system of equations can be
seen as a nonlinear extension of the classical Lyapunov equation for the
multivariate (linear) Ornstein-Uhlenbeck process \citep{Gardiner85}, in which
$W_{i\ell}\:\Sigma_{j\ell}$ would replace $W_{i\ell}\: \Gamma_{j\ell}$ inside
the sum. Such linear equations have been derived previously in the context of
balanced networks of threshold binary units
(\citealp{Renart10,Barrett12,Dahmen16}).

We can also obtain the lagged cross-covariances by integrating the following
differential equation -- obtained using similar methods as above -- over
$s>0$:
\begin{equation}\label{eq:regression}
\frac{\d \Sigma_{ij}(t,s)}{\d s} = \frac1{\tm_j}\left(
-\Sigma_{ij}(t,s) + \sum_\ell W_{j\ell}\: \Gamma_{i\ell}(t,s)
\right)
\end{equation}
Covariances for negative time lags are then obtained from those at positive
lags, since:
\begin{equation}
\b{\Sigma}(t,-s) = \b{\Sigma}(t-s,s)^\trans.
\end{equation}
In the stationary case, we have simply $\b\Sigma(\infty,-s) =
\b\Sigma(\infty,s)^\trans$. \Cref{eq:regression} can be integrated with initial
condition $\b{\Sigma}(t,0)$ which is itself obtained by integrating
\Cref{eq:mean,eq:fpcov}.

%<<<2 Moment closure
\subsection{Calculation of nonlinear moments}

To be able to integrate \Cref{eq:mean,eq:fpcov,eq:regression}, we need to
express the nonlinear moments $\boldsymbol\nu$ and $\b\Gamma$ as a function of
$\boldsymbol\mu$ and $\b\Sigma$. This is a moment closure problem, which in
general cannot be solved exactly. Here, we will approximate $\boldsymbol\nu$
and $\b\Gamma$ by making a Gaussian process assumption for $\b{u}$: we assume
that for any pair of neurons $(i,j)$ and any pair of time points $(t,t+s)$, the
potentials $u_i(t)$ and $u_j(t+s)$ are jointly Gaussian, i.e.\
\begin{equation} \label{eq:gaussianassumption}
  \forall (i,j),\ \forall (t,s)\qquad
\left(\begin{array}{c} u_i(t)\\u_j(t+s)\end{array}\right) \sim
  \mathcal{N}\left[
  \left(\begin{array}{c} \mu_i(t) \\ \mu_j(t+s)\end{array}\right),
  \left(\begin{array}{lr} \Sigma_{ii}(t,0) & \Sigma_{ij}(t,s)\\
                          \Sigma_{ij}(t,s) & \Sigma_{jj}(t+s,0)
        \end{array}\right)
  \right]
\end{equation}
In other words, we systematically and consistently ignore all moments of order
3 or higher.  This is the strongest assumption we make here, but its validity can
always be checked empirically by running stochastic simulations.  For certain
firing rate nonlinearities $f$ (in particular, threshold-power law functions),
the Gaussian process assumption will enable a direct and exact calculation of
$\boldsymbol\nu$ and $\b\Gamma$ given $\boldsymbol\mu$ and $\b\Sigma$, with no
need to linearise the dynamics, as detailed below.

From now on, we will drop the time dependence from the notations to keep the
derivations uncluttered, with the understanding that in using the results that
follow to compute second-order moments such as $\Gamma_{ij}(t,s)$ or
$\Lambda_{ij}(t,s)$, the quantities $\mu_i$ and variances $\Sigma_{ii}$ that
regard neuron $i$ will have to be evaluated at time $t$, those that regard
neuron $j$ evaluated at time $t+s$, and any covariance $\Sigma_{ij}$ will have to
be understood as $\Sigma_{ij}(t,s)$ as defined in \Cref{eq:def:sigma}.
 
Using the Gaussian assumption, mean firing rates become Gaussian integrals:
\begin{equation}\label{eq:integral1}
\nu_i \quad=\quad \int \mathcal{D}z \cdot f\left(\mu_i + z\sig{i}\right)
\end{equation}
where $\mathcal{D}z$ denotes the standard Gaussian measure, and $f(\cdot)$ is
the firing rate nonlinearity (cf.\ \Cref{eq:nonlinearity}). Similarly,
\begin{equation}\label{eq:integral2}
\Gamma_{ij} \quad=\quad \iint \d u_i \: \d u_j \cdot \mathcal{N}\left[
  \left( \begin{array}{c} u_i \\ u_j \end{array} \right);
  \left( \begin{array}{c} \mu_i \\ \mu_j \end{array} \right),
  \left( \begin{array}{cc} \Sigma_{ii} & \Sigma_{ij} \\ 
                           \Sigma_{ij} & \Sigma_{jj} \end{array} \right)
  \right] \cdot (u_i-\mu_i) \cdot f(u_j)
\end{equation}
To calculate $\Gamma_{ij}$, we make use of the fact that the elliptical
Gaussian distribution with correlation $c_{ij}=\Sigma_{ij}/\sqrt{\Sigma_{ii}
\Sigma_{jj}}$ in \Cref{eq:integral2} can be turned into a spherical Gaussian
via a change of variable:
\begin{equation}
\Gamma_{ij} = \iint \mathcal{D}z \: \mathcal{D}z' \cdot
  \sig{i} \left(z' c_{ij} + z \sqrt{1-c_{ij}^2}\right) \cdot 
    f\left(\mu_j + z' \sig{j}\right)
\end{equation}
Now, the integral over $z$ can be performed inside the other one, and clearly
vanishes. We are left with
\begin{equation}\label{eq:integral3}
\Gamma_{ij} = c_{ij}\sig{i} 
  \int \mathcal{D}u \cdot u \cdot f\left(\mu_j + u\sig{j}\right)
\end{equation}
Finally, integrating by part (assuming the relevant integrals exist), we obtain
a simpler form: 
\begin{equation}\label{eq:gamma}
\Gamma_{ij} = \Sigma_{ij} \gamma_j
\end{equation}
with
\begin{equation}\label{eq:littlegamma}
\gamma_j \equiv \int \mathcal{D}u \cdot f'\left(\mu_j + u\sig{j}\right)
\end{equation}

The one-dimensional integrals in \Cref{eq:integral1,eq:littlegamma} turn out to
have closed-form solutions for a number of nonlinearities $u\mapsto f(u)$,
including the exponential $f(u)\propto\exp(u)$ \citep{Buesing12} and the class
of threshold power-law nonlinearities of the form $f(u) = k \lfloor u
\rfloor_+^n$ for any integer exponent $n\geq 1$, which closely match the
behavior of real cortical neurons under realistic noise conditions
\citep{Priebe04,Miller02}.

In the threshold-powerlaw case $r=k\lfloor u \rfloor_+^n$, integration by parts
yields the following recursive formulas:
\begin{align}
\nu_i^{(n)} &= \begin{cases}
\displaystyle k\mu_i\: \psi\left(\frac{\mu_i}{\sig{i}}\right)
    + k\sig{i} \: \phi\left(\frac{\mu_i}{\sig{i}} \right) & {\rm if}\ n=1\\
\displaystyle \mu_i \nu_i^{(1)} + k \Sigma_{ii} \psi\left(\frac{\mu_i}{\sig{i}}\right)
   & {\rm if}\ n=2\\
\displaystyle \mu_i \: \nu_i^{(n-1)} + (n-1) \Sigma_{ii} \nu_i^{(n-2)}
   & {\rm otherwise}
\end{cases}\label{eq:conversionmean}\\
\gamma_i^{(n)} &= \begin{cases}
\displaystyle k \psi\left(\frac{\mu_i}{\sig{i}}\right) & {\rm if}\ n=1\\
\displaystyle n \: \nu_i^{(n-1)} & {\rm otherwise}
\end{cases} \label{eq:conversioncov}
\end{align}
where $\phi$ and $\psi$ denote the standard Gaussian probability density
function and its cumulative density function respectively.
 
Plugging the moment-conversion results of
\Cref{eq:conversionmean,eq:conversioncov} into the equations of motion
(\Cref{eq:mean,eq:fpcov,eq:regression}) yields a system of self-consistent,
\emph{deterministic} differential equations for the temporal evolution of the
potential distribution (or its moments). These equations can be integrated
straightforwardly given the initial moments at time $t=0$. 
       
If we are interested in the stationary distribution of $\b{u}(t)$ or
$\b{r}(t)$, i.e.\ in the case where $h_i(t) = \mathrm{constant}$, we can start
from any valid initial condition $(\boldsymbol\mu,\b\Sigma)$ with
$\b\Sigma\succ 0$, and let the integration of \Cref{eq:mean,eq:fpcov} converge
to a fixed point. In the ergodic case, this will indeed converge to a unique
stationary distribution, independent of the initial conditions used to
integrate the equations of motion. If there are several fixed points, this
procedure yields the moments of the stationary distribution of $\b{u}$
conditioned on the initial conditions, i.e.\ will discover only one of the
fixed points and return the moments of the fluctuations around it. Our approach
cannot capture multistability explicitly, that is, it ignores the possibility
that the network could change its set point with non-zero probability.

Finally, let us emphasise that we have never required that fluctuations be
small.  As long as the Gaussian assumption holds for the membrane potentials
(\Cref{eq:gaussianassumption}), we expect to obtain accurate solutions which is
confirmed below in our numerical examples. This is because we have been able to
express the moments of $\b{r}$ as a function of those of $\b{u}$ in closed
form, \emph{without approximation}.

% ------------------------------------------------------------------------

%<<<2 Temporally correlated noise
\subsection{Extension to temporally correlated input noise}

So far, we have considered spatially correlated, but temporally white, external
noise sources. We now extend our equations to the case of noise with
spatiotemporal correlations, assuming space and time are separable, and
temporal correlations in the input fall off exponentially with time constant
$\tau_\eta$: 
\begin{equation}\label{eq:dynamics-slow-noise}
\tau_i \: \frac{\d u_i}{\d t} =
  -u_i(t) + h_i(t) + \sum_\ell W_{i\ell}\: r_\ell(t)
  + \eta_i(t)
\end{equation}
with $\langle \eta_i(t) \rangle = 0$ and $\langle \eta_i(t) \: \eta_j(t+s)
\rangle \equiv \Sigma_{ij}^\eta e^{-|s|/\tau_\eta}$. The equation for the mean
voltages (\Cref{eq:mean}) does not change. For the voltage covariances,
however, \Cref{eq:fpcov} becomes
\begin{align}
\label{eq:fpcovslow}
\frac{\d\Sigma_{ij}(t,0)}{\d t} \quad&=\quad
 \frac1{\tm_i} \left(-\Sigma_{ij}(t,0)
   + \sum_\ell W_{i\ell} \: \Gamma_{j\ell}(t,0)
   + \Sigma^\star_{ij}(t) \right)\nonumber \\
 &+\quad
 \frac1{\tm_j} \left(-\Sigma_{ij}(t,0)
   + \sum_\ell W_{j\ell} \: \Gamma_{i\ell}(t,0)
   + \Sigma^\star_{ji}(t) \right)
\end{align} 
with the definition $\Sigma^\star_{ij}(t) \equiv \langle \eta_i(t) \:
\s{u}_j(t) \rangle$. These moments can also be obtained by simultaneously
integrating the following:
\begin{align}
\label{eq:fpcovslow2}
\frac{\d \Sigma^\star_{ij}(t)}{\d t} \quad&=\quad 
  -\frac{\Sigma^\star_{ij}(t)}{\tau_\eta} + \frac1{\tm_j} \left( - \Sigma^\star_{ij}(t)
  + \Sigma_{ij}^\eta  
  + \sum_\ell W_{j\ell} \: \Gamma^\star_{i\ell}(t) 
  \right)
\end{align}
with an analogous definition $\Gamma^\star_{ij}(t) \equiv \langle \eta_i(t) \:
\s{r}_j(t) \rangle = \Sigma_{ij}^\star(t) \gamma_j(t)$, which can be expressed
self-consistently as a function of $\mu_j(t)$, $\Sigma_{jj}(t)$, and
$\Sigma^\star_{ij}(t)$ according to \Cref{eq:gamma,eq:conversioncov}.
Altogether, \Cref{eq:mean,eq:fpcovslow,eq:fpcovslow2} form a set of closed and
coupled differential equations for $\{\mu_i(t)\}_i$ and
$\{\Sigma_{ij}(t,0)\}_{i\geq j}$ which can be integrated straightforwardly.

Temporal correlations are given by (for $s>0$): 
\begin{equation}
\frac{\d \Sigma_{ij}(t,s)}{\d s} \quad=\quad
  \frac1{\tm_j} \left[
     -\Sigma_{ij}(t,s) + \sum_\ell W_{j\ell} \: \Gamma_{i\ell}(t,s)
     + \Sigma^\star_{ji}(t) \: \exp\left(-\frac{s}{\tau_\eta} \right)
  \right]
\end{equation}

%<<<2 Summary of equations
\subsection{Summary of equations}

The equations of motion for the moments of $\b{u}$ can be summarised in matrix
form as follows.

\fbox{\parbox{\linewidth}{
{\bfseries Temporally white input noise}\\[0.5em]
\begin{align}
&\textsc{Dynamics} &\d\b{u} \quad&=\quad
  \b{T}^{-1} \left[ -\b{u}(t) + \b{h}(t) + \b{W}\b{r}(t) \right] \d t + d\boldsymbol\chi 
  \label{eq:recap:fast:dyn}\\
&\textsc{Mean} &\frac{\d\boldsymbol\mu(t)}{\d t} \quad&=\quad 
  \b{T}^{-1} \left[ -\boldsymbol\mu(t) + \b{h}(t) + \b{W}\boldsymbol\nu(t) \right]
  \label{eq:recap:fast:mean}\\
&\textsc{Covariance} &\frac{\d\b\Sigma(t,0)}{\d t} \quad&=\quad
  \b\Sigma^\chi + \weff(t)\b\Sigma(t,0) + \b\Sigma(t,0) \weff(t)^\trans 
  \label{eq:recap:fast:cov}\\
&\textsc{Lagged cov.} &\frac{\d\b\Sigma(t,s)}{\d s} \quad&=\quad
  \b\Sigma(t,s) \weff(t+s)^\trans
  \label{eq:recap:fast:lagged}
\end{align}
where we have defined $\b{T} = \diag(\tau_1, \ldots, \tau_N)$, and $\weff(t)
\equiv \b{T}^{-1} \left[\b{W} \diag(\gamma_1(t),\ldots,\gamma_N(t)) -
\b{I}\right]$. In these equations, $\boldsymbol\nu(t)$ and $\boldsymbol\gamma(t)$
are given in closed form as functions of $\boldsymbol\mu(t)$ and $\b\Sigma(t,0)$
according to \Cref{eq:conversionmean,eq:conversioncov}.
}}

\fbox{\parbox{\linewidth}{
{\bfseries Temporally correlated input noise}\\[0.5em]
\begin{align}
&\textsc{Dynamics} &\frac{\d\b{u}}{\d t} \quad&=\quad
  \b{T}^{-1}\left[ -\b{u}(t) + \b{h}(t) + \b{W}\b{r}(t) + \boldsymbol\eta(t)\right]
  \label{eq:recap:slow:dyn}\\
&\textsc{Mean} &\frac{\d\boldsymbol\mu(t)}{\d t} \quad&=\quad 
  \b{T}^{-1} \left[ -\boldsymbol\mu(t) + \b{h}(t) + \b{W}\boldsymbol\nu(t) \right]
  \label{eq:recap:slow:mean}\\
&\textsc{Covariance} &\frac{\d\b\Sigma(t,0)}{\d t} \quad&=\quad
  \left[\b{T}^{-1}\b\Sigma^\star(t)\right] +
  \left[\b{T}^{-1}\b\Sigma^\star(t)\right]^\trans +
  \weff(t)\b\Sigma(t,0) + \b\Sigma(t,0) \weff(t)^\trans 
  \label{eq:recap:slow:cov}\\
& & \frac{\d \b\Sigma^\star(t)}{\d t} \quad&=\quad
  -\frac1{\tau_\eta} \b\Sigma^\star(t) + \b\Sigma^\eta \b{T}^{-1} +
    \b\Sigma^\star(t) \weff(t)^\trans
    \label{eq:recap:slow:covstar}\\
&\textsc{Lagged cov.} &\frac{\d\b\Sigma(t,s)}{\d s} \quad&=\quad
  e^{-s/\tau_\eta} \left[\b{T}^{-1}\b\Sigma^\star(t)\right]^\trans +
  \b\Sigma(t,s) \weff(t+s)^\trans
  \label{eq:recap:slow:lagged}
\end{align}
with the same definitions of $\b{T}$ and $\weff(t)$ as above,
and with $\boldsymbol\nu(t)$ and $\boldsymbol\gamma(t)$ again given by
\Cref{eq:conversionmean,eq:conversioncov} as functions of $\boldsymbol\mu(t)$
and $\boldsymbol\Sigma(t,0)$.
}}

\paragraph{Note on implementation} We favor using the equations of motion in
their matrix form, as we can then use highly efficient libraries for vectorised
operations, especially matrix products (we use the OpenBLAS library). We integrate
\Cref{eq:recap:fast:mean,eq:recap:fast:cov,eq:recap:fast:lagged} and
\Cref{eq:recap:slow:mean,eq:recap:slow:cov,eq:recap:slow:covstar,eq:recap:slow:lagged}
using the classical Euler method with a small time step $\delta_t=0.1$~ms.
When interested in the stationary moments, a good initial condition from which
to start the integration is given by the case $\b{W}=0$, i.e.\
$\boldsymbol\mu(0)\equiv\b{h}$, $\b\Sigma^\star(0) \equiv
\b\Sigma^\eta \diag\left(\frac{1}{1+\boldsymbol\tau/\tau_\eta}\right)$,
and $\b\Sigma(0,0)$ obtained by solving a simple Lyapunov equation.

Care should be taken in integrating the covariance flow of
\Cref{eq:recap:fast:cov,eq:recap:slow:cov} to preserve the positive
definiteness of $\b\Sigma$ at all times. We do this for
\Cref{eq:recap:fast:cov} using the following integrator \citep{Bonnabel12}:
\begin{equation}
\b\Sigma(t+\delta_t,0) = 
\left[\b{I} + \delta_t \weff(t)\right] 
\b\Sigma(t,0) 
\left[\b{I}+\delta_t \weff(t)\right]^\trans
+ \delta_t \b\Sigma^\chi
\end{equation}
and analogously for \Cref{eq:recap:slow:cov}. The complexity is $\mathcal{O}(T
N^3)$ where $T$ is the number of time bins, and $N$ is the number of neurons.
In comparison, stochastic simulations of \Cref{eq:dynamics-fast-noise} cost
$\mathcal{O}(K T N^2)$, where $K$ is a certain number of independent trials
that must be simulated to get an estimate of activity variability. This
complexity is only quadratic in $N$, but in practice $K$ will have to be large
for the moments to be accurately estimated.  Moreover, the generation of random
numbers in Monte-Carlo simulations is expensive. In all the numerical examples
given below, theoretical values were obtained at least 10 times faster than the
corresponding Monte-Carlo estimates, given a decent accuracy criterion
(required number of trials). Where appropriate, one could also apply low-rank
reduction techniques to reduce the complexity of the equations of motion to
$\mathcal{O}(T N^2)$ (e.g.\ in the spirit of \citealp{Bonnabel12}); this is left
for future work.   

%<<<2 Firing rate correlations
\subsection{Firing rate correlations}

So far we have obtained results for pairwise membrane potential covariances
$\Sigma_{ij}(t,s)$, but, as indicated above, these can also be translated into
covariances between the corresponding rate variables,
$\Lambda_{ij}(t,s)\equiv\left\langle \s{r}_i(t) \: \s{r}_j(t+s) \right\rangle$,
which we will need later to compute spike count statistics (e.g.\ Fano factors
or correlations). To shorten the notation, we again drop the time dependence,
with the same understanding that in using the equations that follow to compute
$\Lambda_{ij}(t,s)$, the quantities $\mu_i$ and variances $\Sigma_{ii}$ that
regard neuron $i$ will have to be evaluated at time $t$, those that regard
neuron $j$ evaluated at time $t+s$, and the covariance $\Sigma_{ij}$ will have
to be understood as $\Sigma_{ij}(t,s)$.

\begin{figure}[t]
\centering
\includegraphics[width=0.9\textwidth]{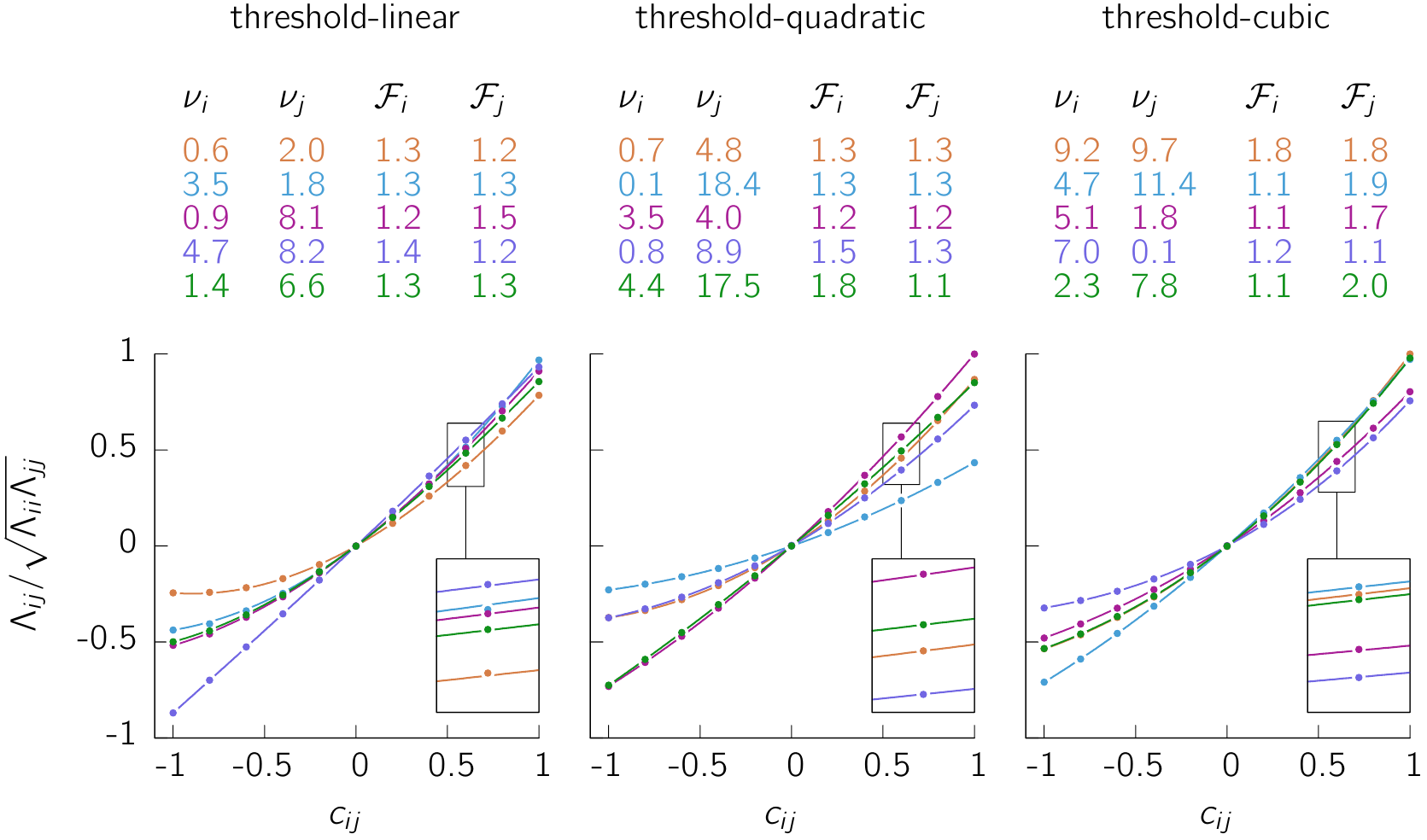}
\caption{\label{fig:ansatz}Numerical validation of our ansatz for rate
covariances.  Shown in 5 different colors are the results of 5 different
simulations in which $\sqrt{\Sigma_{ii}}$ (resp.\ $\sqrt{\Sigma_{jj}}$) was
drawn uniformly between $1$ and $4$, and $\mu_i$ (resp.\ $\mu_j$) was chosen so
as to achieve a mean firing rate $\nu_i$ (resp.\ $\nu_j$) drawn from a Gamma distribution
with a mean $5$~Hz and a shape parameter of $1$. Dots show $\Lambda_{ij}$ as a
function of $c_{ij}$, as estimated via Monte-Carlo integration of
\Cref{eq:lambda2d} with a million samples. Solid lines show \Cref{eq:ratecov},
which is by construction always exact at $c_{ij}=\pm 1$, and always has the
right slope at $c_{ij}=0$.  The corresponding tables show the mean firing rates
and the Fano factors of both neurons (Fano factors were computed using
\Cref{eq:fanoapprox} with $\tau_{\rm A}=50$~ms and $T=100$~ms).  The figures
indicate that our approximation is highly accurate over a range of parameters
corresponding to physiological values of firing rates and Fano factors. {\it
Parameters: $k=3$ (threshold-linear), $k=0.3$ (threshold-quadratic), $k=0.02$
(threshold-cubic).}}
\end{figure}

Under the same Gaussian process assumption as above, we have
\begin{equation}
\Lambda_{ij} \quad=\quad \iint \d u \: \d u' \cdot \mathcal{N}\left[
  \left( \begin{array}{c} u \\ u' \end{array} \right);
  \left( \begin{array}{c} \mu_i \\ \mu_j \end{array} \right),
  \left( \begin{array}{cc} \Sigma_{ii} & \Sigma_{ij} \\ 
                           \Sigma_{ij} & \Sigma_{jj} \end{array} \right)
  \right]  \cdot (f(u) - \nu_i) \cdot (f(u')-\nu_j)
\label{eq:lambda2d}
\end{equation}
Calculating this double integral exactly seems infeasible.  However, as
detailed below, we were able to derive an analytical approximation that is
highly accurate over a broad range of physiologically relevant values
(\Cref{fig:ansatz}), for the class of threshold power-law nonlinearities $f(u)
= k\lfloor u \rfloor_+^n$.

Numerical explorations of the behaviour of $\Lambda_{ij}$ as a function of the
moments of $u_i$ and $u_j$ suggest the following \emph{ansatz}:
\begin{equation}\label{eq:ratecov}
\Lambda_{ij} \quad=\quad 
  \alpha_{ij}^{(3)} c_{ij}^3 + \alpha_{ij}^{(2)} c_{ij}^2 
  + \alpha_{ij}^{(1)} c_{ij} 
\end{equation}
where $c_{ij} = \Sigma_{ij}/\sqrt{\Sigma_{ii}\Sigma_{jj}}$ is the correlation
coefficient between $u_i$ and $u_j$ (at time $t$ and lag $s$), and the three
coefficients $\alpha_{ij}^{(\cdot)}$ do not depend on $c_{ij}$ (though they
depend on the marginals over $u_i$ and $u_j$, as detailed below) and can be
computed exactly. We focus on $\Lambda_{ij}$ as a function of $c_{ij}$ -- we
abuse the notation of \Cref{eq:def:lambda} and write this dependence as
$\Lambda_{ij}(c_{ij})$.  Clearly, $\Lambda_{ij}(0) = 0$. Next, we note that  
\begin{subequations}
\label{eq:ansatzconstraints}
\begin{align}
\alpha_{ij}^{(2)} \quad&=\quad  
   \left[\Lambda_{ij}(+1) + \Lambda_{ij}(-1)\right]/2\\
\alpha_{ij}^{(1)}+\alpha_{ij}^{(3)} \quad&=\quad  
   \left[\Lambda_{ij}(+1) - \Lambda_{ij}(-1)\right]/2\\
\alpha_{ij}^{(1)} \quad&=\quad
   \left. \frac{\d \Lambda_{ij}(c_{ij})}{\d c_{ij}} \right|_{c_{ij}=0}\label{eq:alpha1}
\end{align}
\end{subequations}
where, specialising to the threshold-power law nonlinearity $f(u)=k\lfloor u
\rfloor_+^n$,
\begin{equation}\label{eq:alphapm}
\Lambda_{ij}({\color{red}{\pm}} 1) \quad=\quad
  - \nu_i \: \nu_j
  + k^2 \int \mathcal{D}u\
  \lfloor \mu_i {\color{red}\pm} u \sig{i} \rfloor_+^n  \
  \lfloor \mu_j + u \sig{j}\rfloor_+^n
\end{equation}
After some algebra, \Cref{eq:alpha1} yields
\begin{equation}
\alpha_{ij}^{(1)} = \left( \gamma_i \sqrt{\Sigma_{ii}} \right)
  \left( \gamma_j \sqrt{\Sigma_{jj}} \right)
\end{equation}
where $\Gamma_{ii}$ and $\Gamma_{jj}$ were derived previously in
\Cref{eq:gamma,eq:conversioncov}. To compute $\Lambda_{ij}(+1)$, let us define
more generally
\begin{equation}
A_{ij}^{(n,m)} \equiv k^2 \int \mathcal{D}u \ 
  \lfloor \mu_i + u\sig{i} \rfloor_+^n \
  \lfloor \mu_j + u\sig{j} \rfloor_+^m
\end{equation}
keeping in mind that we are ultimately interested in $A_{ij}^{(n,n)}$, since
$\Lambda_{ij}(+1) = -\nu_i \nu_j + A_{ij}^{(n,n)}$. In the
following, we assume that $\mu_i/\sig{i} \geq \mu_j \sig{j}$. If the opposite
holds, then the indices $i$ and $j$ \emph{must be swapped at this stage}
(this is just a matter of notation). Using techniques similar to the integral
calculations carried out previously (mostly, integration by parts), we derived
the following recursive formula valid for $0 \leq n \leq m$:
\begin{equation}\label{eq:aijrecursion}
A_{ij}^{(n,m)} = \begin{cases}
\displaystyle  k\nu_j^{(m)} & {\rm if}\ n=0\\
\displaystyle  k \mu_i \nu_j^{(m)} + k\sqrt{\frac{\Sigma_{ii}}{\Sigma_{jj}}} \:
  \Gamma_{jj}^{(m)} & {\rm if}\ n=1\\
\displaystyle  \mu_i A_{ij}^{(n-1,m)} + (n-1)\Sigma_{ii} A_{ij}^{(n-2,m)}
  + m\sqrt{\Sigma_{ii}\Sigma_{jj}} A_{ij}^{(n-1,m-1)} & {\rm otherwise}
  \end{cases}
\end{equation}
where $\nu_j^{(m)}$ and $\Gamma_j^{(m)}$ were calculated previously (cf.\
\Cref{eq:conversionmean,eq:conversioncov}). 

The calculation of $\Lambda_{ij}(-1)$ is slightly more involved, but follows a
similar logic. This time we assume that $\mu_i/\sig{i} > -\mu_j/\sig{j}$
(otherwise, $\Lambda_{ij}(-1)=0$), and we define
\begin{equation}
B_{ij}^{(n,m)} \equiv k^2 \int \mathcal{D}u \ 
  \lfloor \mu_i - u\sig{i} \rfloor_+^n \
  \lfloor \mu_j + u\sig{j} \rfloor_+^m
\end{equation} 
Similar to \Cref{eq:aijrecursion}, we have the following recursive formula for
$1<n\leq m$:
\begin{equation}\label{eq:bijrecursion}
B_{ij}^{(n,m)} = \mu_i B_{ij}^{(n-1,m)} + (n-1)\Sigma_{ii} B_{ij}^{(n-2,m)}
  - m\sqrt{\Sigma_{ii}\Sigma_{jj}} B_{ij}^{(n-1,m-1)}
\end{equation}
However, now the boundary conditions must also be computed recursively:
\begin{align}
B_{ij}^{(0,m)} &= \begin{cases}
  k^2 \left[ \psi_i + \psi_j - 1 \right] & {\rm if}\ m=0\\
  \mu_j B_{ij}^{0,0} + k^2 \sig{j} \left( \phi_j - \phi_i \right)
    & {\rm if}\ m=1\\
  \mu_j B_{ij}^{(0,m-1)} - \sig{j} \left[
    k^2 \phi_i R_{ij}^{m-1} - (m-1)\sig{j} B_{ij}^{(0,m-2)}
  \right] & {\rm otherwise} 
  \end{cases}\\
B_{ij}^{(1,m)} &= \mu_i B_{ij}^{(0,m)} + \sig{i} \begin{cases}
  k^2 \left(\phi_i - \phi_j\right) & {\rm if}\ m=0\\
  k^2 \phi_i R_{ij}^m - m \sig{j} B_{ij}^{(0,m-1)} & {\rm otherwise}
  \end{cases}
\end{align}
where we have used the shorthands $\phi_\ell \equiv \phi(\mu_\ell/\sig{\ell})$,
$\psi_\ell \equiv \psi(\mu_\ell/\sig{\ell})$, and $R_{ij} \equiv \mu_j +
\mu_i\sqrt{\Sigma_{jj}/\Sigma_{ii}}$.

Thus, $\Lambda_{ij}(c_{ij})$ is approximated in closed-form by a third-order
polynomial in $c_{ij}$, given $\mu_i$, $\mu_j$, $\Sigma_{ii}$, $\Sigma_{jj}$ and
$\Sigma_{ij}$. Our polynomial approximation is very accurate over a broad,
physiologically relevant range of parameters (and in fact, even beyond that), for
power law exponents in the physiological range (\Cref{fig:ansatz}). 

In combination with the results of \Cref{sec:umoments}, we can then obtain the
moments of $\b{r}$, which we use in the following section to compute Fano
factors and spike count correlations.

%<<<2 Spike count statistics
\subsection{\label{sec:fano}Spike count statistics}

Under the assumption that each neuron $i$ emits spikes according to an
inhomogeneous Poisson process with rate function $r_i(t)$ (``doubly
stochastic'', or ``Cox'' process)\footnote{This does not affect the form of the
network dynamics which remain rate-based (\Cref{eq:dynamics-fast-noise}); that
is, spikes are generated on top of the firing rate fluctuations given by the
rate model.}, we can compute the joint statistics of the spike counts in some
time window, which is what electrophysiologists often report. Let $C_i^T(t)$
denote the number of spikes that are emitted by neuron $i$ in a window of
duration $T$ starting at time $t$, and let $\kappa_i^T(t) = \int_0^T \d t\
r_i(t)$ be the expected number of spikes in that window, for a given trial and
\emph{given} the underlying rate trace $r_i(t)$. The Fano factor of the
distribution of $C_i^T(t)$ is given by
\begin{equation}\label{eq:fano}
\mathcal{F}_i^T(t) \quad=\quad \frac{{\rm var}\left[C_i^T(t)\right]}
  {{\rm mean}\left[C_i^T(t)\right]}
  \quad = \quad 
  1 + \frac{{\rm var}\left[\kappa_i^T(t)\right]}{\langle \kappa_i^T(t) \rangle}
\end{equation}
with
\begin{equation}
\langle \kappa_i^T(t) \rangle = \int_0^T \d t' \: \nu_i(t+t')
\end{equation}
and
\begin{equation}
{\rm var}\left[\kappa_i^T(t)\right] =
\int_0^T \d s \int_0^T \d s' \: \Lambda_{ii}(t+s,s'-s)
\end{equation}
All expected values such as $\nu_i$ and $\Lambda_{ii}$ have been calculated in
previous sections. We approximate the integrals above by simple Riemann sums
with a discretisation step of $1$~ms. 

\begin{figure}
\centering
\includegraphics[width=0.75\textwidth]{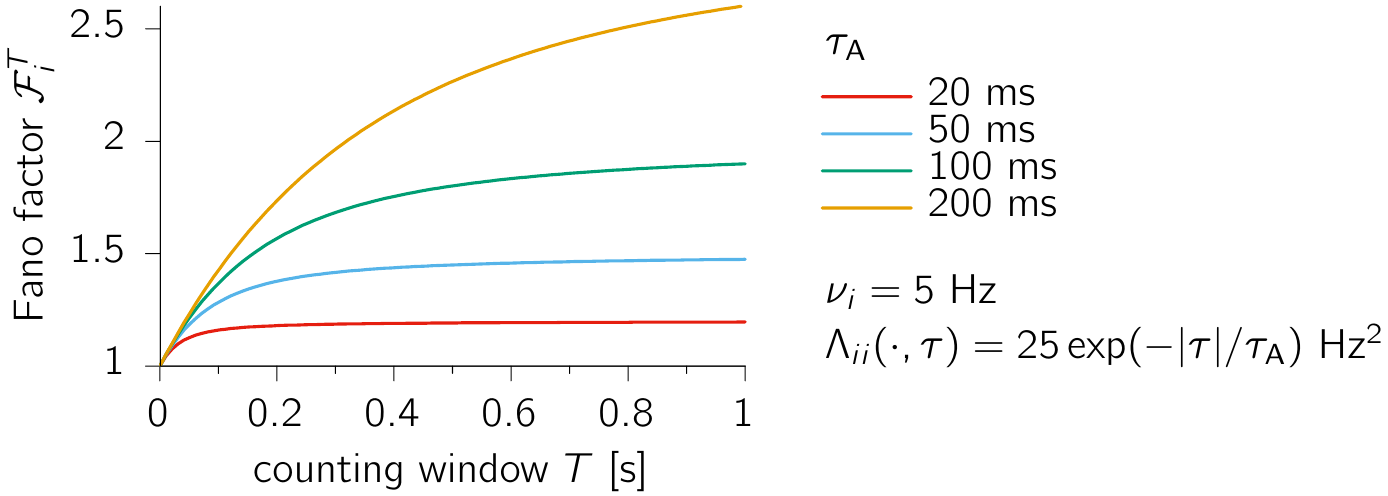}
\caption{\label{fig:fanoapprox}Approximation of the Fano factor when the
autocorrelation is a Laplacian with time constant $\tau_{\sf A}$ (varied here;
cf.\ color legend) -- see also \Cref{eq:fanoapprox}.} 
\end{figure}

In the special case of constant input $\b{h}$ leading to a stationary rate
distribution, \Cref{eq:fano} simplifies to 
\begin{equation}\label{eq:fanostationaryintegral}
\mathcal{F}_i^T = 1+\frac1{T\nu_i}\:
\int_0^T \d s \int_{0}^{T} \d s' \: \Lambda_{ii}(\cdot,s'-s)
\end{equation} 
(the rate variance $\Lambda_{ii}(t,\tau)$ no longer depends on $t$, hence
the notation $\Lambda_{ii}(\cdot,\tau)$). A closed-form approximation can be
derived if the rate autocorrelation is well approximated by a Laplacian with
decay time constant $\tau_{\rm A}$, i.e.\ $\Lambda_{ii}(\cdot,\tau) \approx
\Lambda_{ii}(\cdot,0)\: \exp\left(-|\tau|/\tau_{\rm A}\right)$. In this case,
\Cref{eq:fanostationaryintegral} evaluates to
\begin{equation}\label{eq:fanoapprox}
\mathcal{F}_i^T \quad\approx\quad 1+
  \frac{2\:\tau_{\rm A}\:\Lambda_{ii}(\cdot,0)}{\nu_i}
\left[ 1-\frac{\tau_{\rm A}}T \: 
  \left( 1-e^{-T/\tau_{\rm A}} \right)
\right]
\end{equation}
This expression is shown in \Cref{fig:fanoapprox} as a function of the counting window
$T$, for various autocorrelation lengths $\tau_{\sf A}$.

\begin{figure}[t]
\centering
\includegraphics[width=\textwidth]{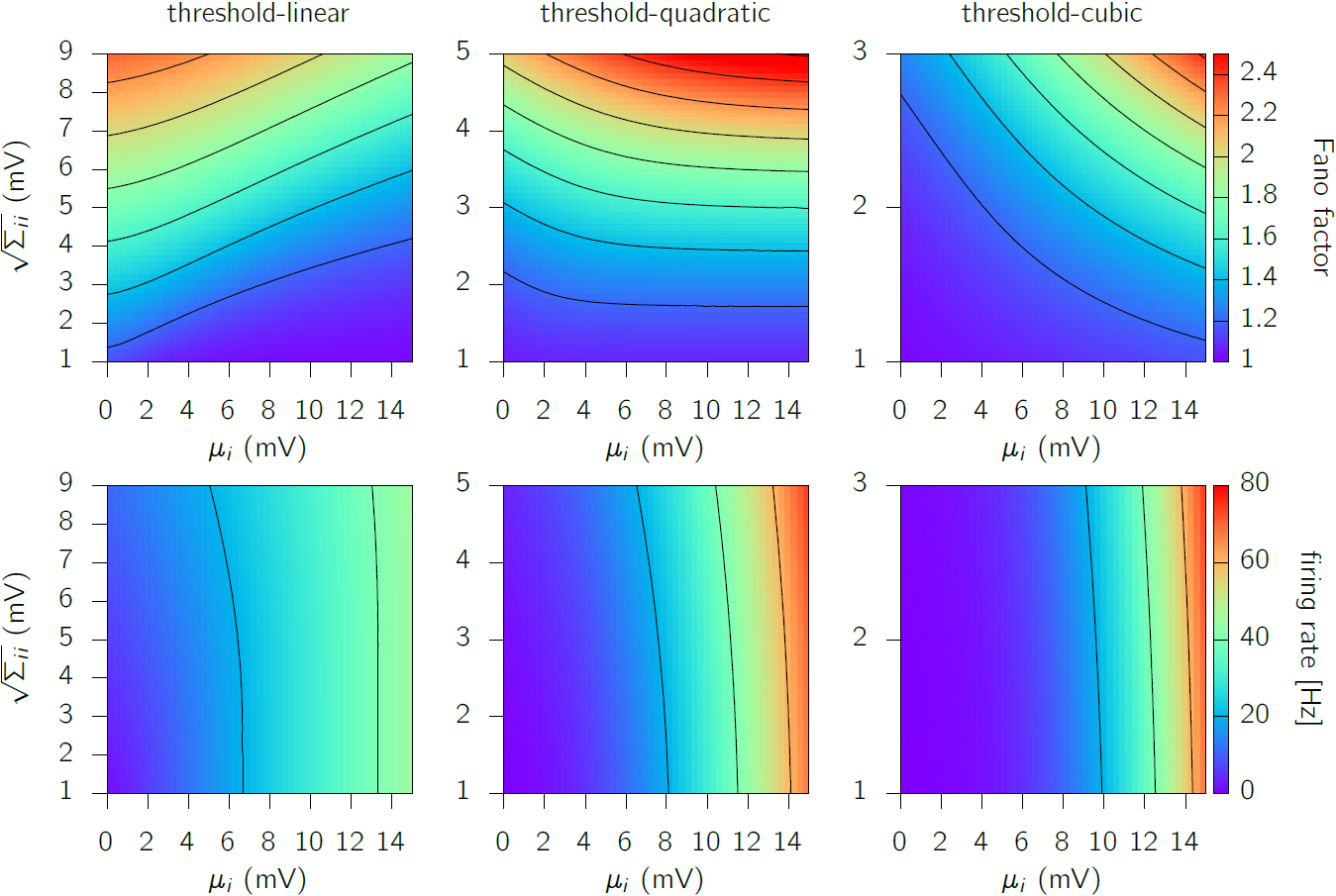}
\caption{\label{fig:fano}Fano factor $\mathcal{F}_i$ (top row) and mean firing
rate $\nu_i$ (bottom row) as a function of membrane potential mean $\mu_i$ and
standard deviation $\sig{i}$, for $n=1$ to $n=3$ (left to right). To calculate
$\mathcal{F}_i$, we used \Cref{eq:fanoapprox}, assuming an autocorrelation time
constant of $\tau_{\rm A}=50$~ms. The multiplicative rescaling $k$ of the
threshold power-law gain function was adjusted so that both Fano factors and
firing rates span roughly the same range of values across $n=1$ to $n=3$, but
the precise value of $k$ has no impact on the shape of the contour lines.}
\end{figure}

The behaviour of $\mathcal{F}_i$ as a function of membrane potential sufficient
statistics $\mu_i$ and $\sig{i}$ is depicted in \Cref{fig:fano}, for various
exponents of the threshold power-law gain function.  In all cases (i.e.\
independent of the exponent $n$), the Fano factor grows with increasing
potential variance. However, the dependence on $\mu_i$ strongly depends on $n$.
For $n=1$, the Fano factor decreases with $\mu_i$, while for $n=3$ it increases.
For $n=2$, the Fano factor has only a weak dependence on $\mu_i$. This
behaviour can be understood qualitatively by linearising the gain function
around the mean potential, which gives us an idea of how the super-Poisson part
of the Fano factor depends on membrane potential statistics:
\begin{equation}\label{eq:si:fanoapproxlinear}
\mathcal{F}_i^T - 1 \propto n \mu_i^{n-2} \Sigma_{ii}
\end{equation}
Therefore, the iso-Fano-factor lines in \Cref{fig:fano} are expected to
approximately obey $\sig{i} \approx \mu_i^{2-n}$, which is roughly what we see
from \Cref{fig:fano}.
 
It is instructive to plug in some typical numbers: assuming a mean firing
rate of $\nu_i= 5$~Hz, a counting window $T=50$~ms, a Fano factor
$\mathcal{F}_i^T=1.5$ (characteristic of spontaneous activity in the cortex,
\citealp{Churchland10}), and an autocorrelation time constant $\tau_{\rm
A}=40$~ms \citep{Kenet03,Berkes11}, then \Cref{eq:fanoapprox} tells us that the
fluctuations in $r_i$ must have a standard deviation of $\sqrt{\Lambda_{ii}}
\simeq 8.5$~Hz.  This is larger than the assumed mean firing rate ($5$~Hz),
thus implying that the underlying rate variable in physiological conditions is
often going to be zero, in turn implying that the membrane potential will often
be found below the threshold of the firing rate nonlinearity $f(u)$.  It is
precisely in this regime that it becomes important to perform the nonlinear
conversion of the moments of $\b{u}$ into those of $\b{r}$, taking into account
the specific form of the nonlinearity; in that same regime, linearisation of
the dynamics of \Cref{eq:dynamics-fast-noise} can become very inaccurate.

Similar derivations can be done for spike count correlations:
\begin{align}
\rho_{ij}^T(t) \quad&\equiv\quad \frac{{\rm cov}\left[C_i^T(t), C_j^T(t)\right]}
  {\sqrt{{\rm var}\left[\vphantom{C_j^T}C_i^T(t)\right]\ 
  {\rm var} \left[C_j^T(t)\right]}}\nonumber\\
  &=\quad \frac{\langle \kappa_i^T(t) \kappa_j^T(t) \rangle
   - \langle \kappa_i^T(t) \rangle \langle \kappa_j^T(t) \rangle}
  {\sqrt{ \left( 
   {\rm var}\left[\vphantom{\kappa_j^T} \kappa_i^T(t) \right] 
     + \langle \kappa_i^T(t) \rangle \right)
   \left( {\rm var}\left[ \kappa_j^T(t) \right] + \langle \kappa_j^T(t) \rangle
   \right)}}\nonumber\\
  &=\quad
  \frac{\displaystyle
   \iint_{[0:T]^2} \d s\: \d s' \ \Lambda_{ij}(t,s'-s)
  }{\sqrt{\left({\rm var}\left[ \vphantom{\kappa_j^T}\kappa_i^T(t) \right] 
    + \langle \kappa_i^T(t) \rangle \right) \left(
   {\rm var}\left[ \kappa_j^T(t) \right] + \langle \kappa_j^T(t) \rangle \right)}}\nonumber\\
  &=\left[\langle \kappa_i^T(t)\rangle \: \langle \kappa_j^T(t) \rangle
  \: \mathcal{F}_i^T(t) \: \mathcal{F}_i^T(t)\right]^{-\frac12}
   \iint_{[0:T]^2} \d s \: \d s' \ \Lambda_{ij}(t,s'-s) \label{eq:countcorr}
\end{align}

In the stationary case, \Cref{eq:countcorr} becomes
\begin{equation}
\rho_{ij}^T \quad=\quad \frac{\displaystyle
  \iint_{[0:T]^2} \d s \: \d s' \: \Lambda_{ij}(\cdot,s'-s)}{
  T\: \sqrt{ \nu_i \nu_j \mathcal{F}_i^T \mathcal{F}_j^T}}
\end{equation}

%<<<1 Numerical validation on random SSN
\section{\label{sec:numerics}Numerical validation}

In this section, we demonstrate the validity of the equations derived in
\Cref{sec:theory} on two examples: a random E/I network with weak and sparse
connections, and a strongly connected, inhibition-stabilized E/I network.

\subsection{Weakly connected random network}

\begin{figure}[b!]
\centering
\includegraphics[width=0.6\textwidth]{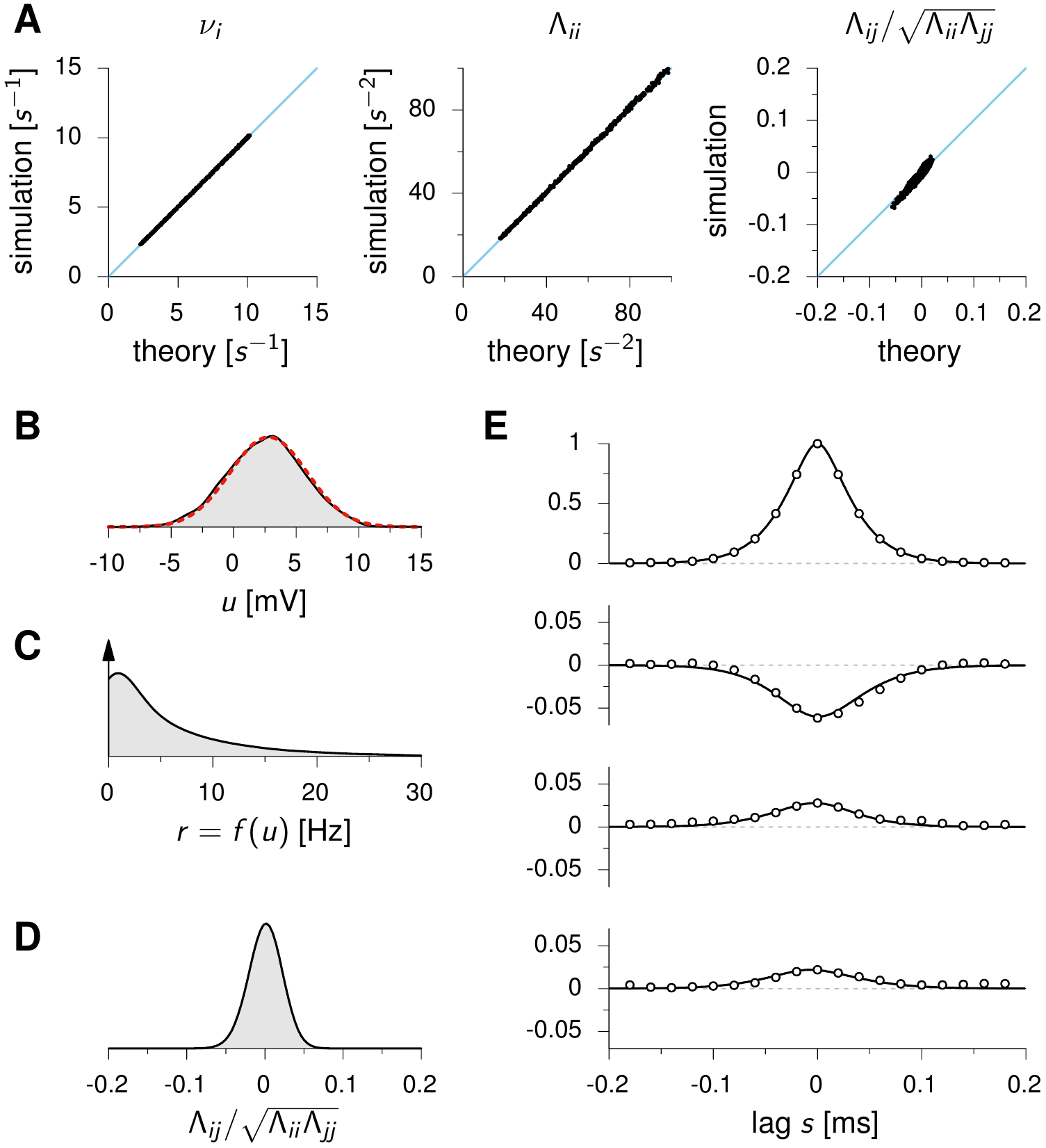}
\caption{\label{fig:random_net}{\bfseries Validation of our theoretical results
on a random, weakly connected E/I network}.
{\bfseries (A)}~Mean firing rates $\{\nu_i\}$, firing rate variances
$\{\Lambda_{ii}\}$, and pairwise firing rate correlations, as predicted
semi-analytically by the moment dynamics equations of \Cref{sec:theory}
(x-axes) and compared to empirical estimates obtained from long stochastic
simulations of \Cref{eq:dynamics-slow-noise} (y-axes).
{\bfseries (B)}~Distribution of membrane potentials $u_i(t)$ for a randomly
chosen neuron (gray), and its Gaussian approximation $\mathcal{N}(u; \mu_i,
\Sigma_{ii})$ used in our theory.
{\bfseries (C)}~Corresponding firing rate distribution of the same neuron as in 
(B).
{\bfseries (D)}~Distribution of pairwise firing rate correlations across the network.
{\bfseries (E)}~Example membrane potential normalized auto- (top row) and
cross- (bottom rows) correlograms
($\Sigma_{ij}(\infty,s)/\sqrt{\Sigma_{ii}\Sigma_{jj}}$), as predicted by the
theory (solid) and estimated from simulations (dots). 
}
\end{figure}
 
The network is made of $N_\e=250$ excitatory and $N_\i=250$ inhibitory neurons,
each with a threshold-quadratic I/O nonlinearity: $f(u) = 0.3\lfloor u
\rfloor_+^2$. We set all intrinsic time constants to $\tau_i = \tau \equiv 20$~ms. The
input noise has no spatial, but only temporal, correlations:
\begin{equation}
\langle \eta_i(t)\eta_j(t+s)\rangle = \frac{\sigma_0^2}{1+\tau/\tau_\eta} \:
  \exp\left(-\frac{|s|}{\tau_\eta}\right) \delta_{ij}
\end{equation}
with $\tau_\eta=50$~ms and $\sigma_0=3$~mV representing the standard deviation
of the membrane potentials if the recurrent connectivity were removed.
Synaptic weights are drawn randomly from the following distribution:
\begin{equation}
W_{ij} = \frac{\alpha s_j}{\sqrt{N}} \times \left\{ \begin{array}{rl}
  1 & \text{with probability 0.2}\\
  0 & \text{otherwise}
\end{array} \right.
\end{equation}
where $\alpha$ is a global scaling factor (see below), $s_j$ is a presynaptic
signed factor equal to $+1$ if neuron $j$ is excitatory ($1\leq j \leq N/2$),
and equal to $-\gamma$ if neuron $j$ is inhibitory ($N/2 < j \leq N$). We set
$\gamma=3$ to place the network in the inhibition-dominated regime where the
average pairwise correlation among neurons is weak \citep{Renart10,
Hennequin12}. The scaling factor $\alpha = 2.2$ was chosen such that the
network is effectively weakly connected and thus far from instability.

The network is fed with a constant input $\b{h} = \b{u^\star} -
\b{W}f(\b{u}^\star)$, such that, in the absence of stochasticity, the network
would have a (stable) fixed point at $\b{u}=\b{u}^\star$ -- we drew the
elements of $\b{u}^\star$ from a uniform distribution between $1$ and $4$~mV.

We simulated the stochastic dynamics of the network
(\Cref{eq:dynamics-slow-noise}) for $5000$ seconds, and computed empirical
estimates of the moments in the steady-state, stationary regime (mean firing
rates, firing rate variances, firing rate correlations, and full membrane
potential cross-correlograms). We found those estimates to agree very well with
the semi-analytical solutions obtained by integrating the relevant equations of
motion derived in \Cref{sec:theory} under the Gaussian assumption
(\Cref{fig:random_net}A).  Membrane potentials are indeed roughly normally
distributed (\Cref{fig:random_net}B), with a standard deviation on the same
order as the mean, yielding very skewed distributions of firing rates
(\Cref{fig:random_net}C). Due to the effectively weak connectivity, pairwise
correlations among firing rate fluctuations are weak in magnitude
(\Cref{fig:random_net}D). Membrane potential cross-correlograms have a simple,
near-symmetric structure (random, weakly connected networks are close to
equilibrium) and are well captured by the theory.

\subsection{Strongly connected, inhibition-stabilized random network}

We now consider a strongly connected E/I network much further away from
equilibrium than the weakly-connected network of \Cref{fig:random_net}. This
network operates in the inhibition-stabilized regime, whereby excitatory
feedback is strongly distabilizing on its own, but is dynamically stabilized by
feedback inhibition. The details of how we obtained such a network are largely
irrelevant here (but see \citealp{Hennequin14}). All the parameters of the network
will be posted online together with the code to ensure reproducibility. 

\begin{figure}
\centering
\includegraphics[width=0.6\textwidth]{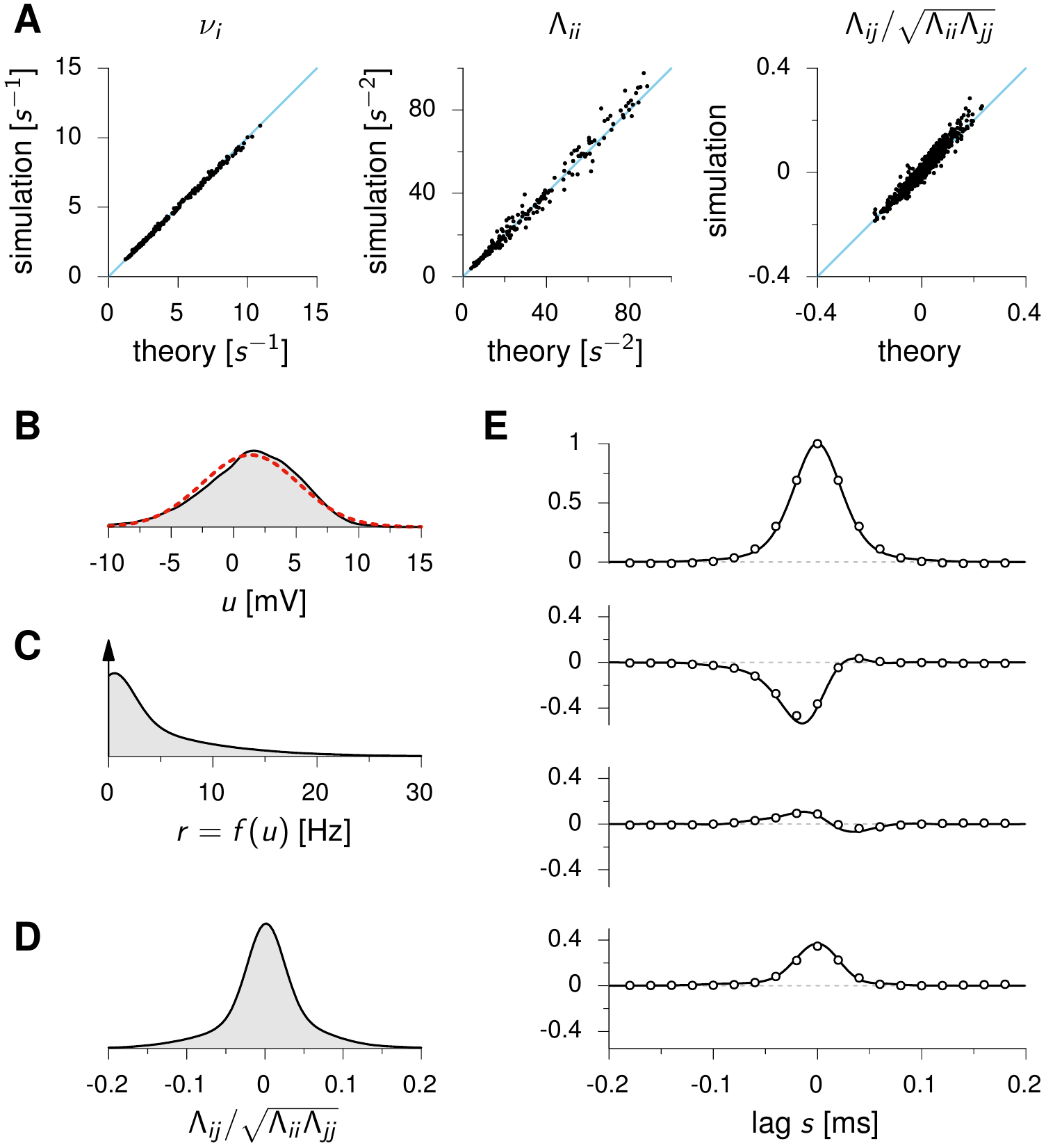}
\caption{\label{fig:random_isn}{\bfseries Validation of our theoretical results
on a random, strongly connected, inhibition-stabilized E/I network}. 
The figure follows the same format as in \Cref{fig:random_net} above; see
caption there.  }
\end{figure}

Our results are summarized in \Cref{fig:random_isn}, in exactly the same format
as in \Cref{fig:random_net}. This network strongly amplifies the input noise
process along a few specific directions in state space, leading to strong
(negative and positive) pairwise correlations among neuronal firing rates
\Cref{fig:random_isn}D. Thus, the central limit theorem -- which would in
principle justify our core assumption that membrane potentials are jointly
Gaussian, as large (and low-pass-filtered) sums of uncorrelated variables --
may not apply.  Indeed, most membrane potential distributions have substantial
(negative) skewness (\Cref{fig:random_isn}B), leading to some inaccuracies in
our semi-analytical calculation of the moments (\Cref{fig:random_isn}A), which
we find are reasonably small nonetheless.  We leave the extension of our theory
to third-order moments (e.g.\ \citealp{Dahmen16}) for future work. Finally, the
strong connectivity gives rise to non-trivial temporal structure in the joint
activity of pairs of neurons reflecting non-equilibrium dynamics.  Even in this
regime, our results capture the membrane potential cross-correlograms well
(\Cref{fig:random_isn}E).

%<<<1 Acknowledgment
\section*{Acknowledgments}
This work was supported by the Swiss National Science Foundation (GH)
and the Wellcome Trust (GH, ML).
 
\clearpage
%<<<1 Bibliography

\bibliographystyle{apalike}
\bibliography{mybib.bib}

\end{document}